\documentclass[aps, prd,showpacs, superscriptaddress, groupedaddress]{revtex4-2} 
\usepackage{graphicx}  
\usepackage{subcaption}
\usepackage{float}
\usepackage{tabularx}
\usepackage{array}
\usepackage{amssymb}
\usepackage{mathtools}
\usepackage{dcolumn}
\usepackage{color}
\usepackage[pagebackref=false, colorlinks=true]{hyperref}
\hypersetup{
linkcolor=blue,     
citecolor=blue,     
urlcolor=blue}   
\usepackage{amsmath}
\usepackage{ulem}
\usepackage[utf8]{inputenc}

\begin{document}
\title{Generalized JMN Naked Singularity Models}

\author{Jay Verma Trivedi}
\email{jay.verma2210@gmail.com}
\affiliation{International Centre for Space and Cosmology, School of Arts and Sciences, Ahmedabad University, Ahmedabad-380009 (Guj), India.}
\author{Pankaj S. Joshi}
\email{pankaj.joshi@ahduni.edu.in}
\affiliation{International Centre for Space and Cosmology, School of Arts and Sciences, Ahmedabad University, Ahmedabad-380009 (Guj), India.}

\date{\today}
\begin{abstract}
We construct here a generalization of Joshi-Malafarina-Narayan (JMN) naked singularity spacetimes that arise as equilibrium end states of gravitational collapse with non-vanishing tangential pressure. The generalization introduces density inhomogeneity through a radially dependent mass function $F(r)=(M_0+M_n r^n)r^3$, leading to a two–parameter family of solutions matched smoothly to an exterior Schwarzschild spacetime. 
The observational properties of the spacetime are then examined through shadow formation and emission from accreting matter. We find that in this more general case, as well, when the photon sphere lies in the exterior Schwarzschild region, the shadow is identical to that of a Schwarzschild black hole. Accretion spectra show enhanced high–frequency emission compared to Schwarzschild, while deviations from the original JMN model remain small due to strong constraints on the inhomogeneity parameter. These results indicate that the generalized model effectively serves as a small perturbation of the JMN spacetime, demonstrating the robustness of the physical properties of the original JMN-type naked-singularity geometries.\\
\vspace{0.5cm}
\\
\textbf{keywords}: JMN, naked singularity, gravitational collapse, generalization, shadow, accretion disk.
\end{abstract}
\maketitle

\section{Introduction}
The Cosmic Censorship Conjecture (CCC), originally proposed by Penrose in 1969 \cite{Penrose:1969pc, Hawking:1979ig}, remains one of the central unresolved problems in classical general relativity. Although it plays a fundamental role in our understanding of black hole physics, neither a precise mathematical formulation nor a general proof of the conjecture has been established to date. The singularity theorems of Hawking and Penrose guarantee the formation of spacetime singularities under broad, physically reasonable conditions \cite{HawkingEllis1973}; however, they remain silent on the crucial question of whether such singularities are necessarily hidden behind event horizons or can be visible to distant observers.

Models of gravitational collapse have, however, provided significant insight into this problem. A large body of work over the past several decades has demonstrated that the end state of collapse may be either a black hole or a naked singularity, depending sensitively on the initial conditions of the collapsing matter \cite{Dwivedi:1992fh, Joshi:1993zg, Joshi:1994br, Joshi:2001xi, Hellaby:1985zz, Goncalves:2001pf, Giambo:2002xc, Harada:2001nj, Ori:1989ps, Gundlach:1999cu, Nolan:2002zd, Giacomazzo:2011cv, Ortiz:2011jw, Banerjee:2002sy, Barausse:2010ka, Mosani:2023vtr}. These results indicate that naked singularities arise naturally in general-relativistic gravitational collapse, and that the formation of an event horizon is not enforced solely by the Einstein field equations. As a consequence, CCC remains unproven, and within physically realistic collapse scenarios, it appears to be theoretically violated.

If naked singularities are indeed a generic outcome of gravitational collapse, then CCC ultimately becomes an observational question. In this context, recent studies have focused on identifying potential astrophysical signatures that can distinguish naked singularities from black holes. In particular, analyses of the first image of Sagittarius~A* (Sgr~A*) have highlighted the JMN-1 naked singularity spacetime \cite{Joshi:2011zm}, which arises as an equilibrium configuration of anisotropic fluid collapse with vanishing radial and non-zero tangential pressure, as a viable black hole mimicker \cite{EventHorizonTelescope:2022xqj}.

It is well known that a spherically symmetric distribution of non-interacting particles, in the absence of pressure, cannot remain in equilibrium against its own gravitational attraction and must undergo complete gravitational collapse. The final outcome of such a dust collapse is either a black hole or a naked singularity, depending on the initial density and velocity profiles of the collapsing matter \cite{Joshi:1993zg, WaughLake1988, Newman1986, Christodoulou1984, EardleySmarr1979}. Over the past decades, extensive investigations of gravitational collapse in a wide range of physical settings—such as self-similar models \cite{OriPiran1987, Ori:1989ps, FoglizzoHenriksen1993}, scalar fields \cite{Christodoulou1994, Giambo2005, BhattacharyaGoswamiJoshi2011}, perfect fluids \cite{Harada1998, HaradaMaeda2001, GoswamiJoshi2002, GiamboGiannoniMagliPiccione2004, VillasdaRochaWang2000}, and more general matter configurations—have shown that both black holes and naked singularities naturally arise as possible end states within general relativity \cite{ShapiroTeukolsky1991, ShapiroTeukolsky1992, Lake1991, JoshiKrolak1996, Joshi:2001xi, GoswamiJoshi2007}. 

When pressures are present within the collapsing matter cloud, a scenario that is physically more realistic than dust, complete gravitational collapse is no longer the only possible outcome. An important question then arises: whether and under what conditions equilibrium configurations can emerge dynamically from gravitational collapse. This issue is directly physical, as numerous astrophysical objects, including stars, planets, and galaxies, form through collapse and subsequently settle into long-lived equilibrium states.

In \cite{Joshi:2011zm}, it was demonstrated that the gravitational collapse of a matter cloud with non-vanishing tangential pressure and vanishing radial pressure, starting from regular initial data, can asymptotically approach a variety of equilibrium configurations. While this choice represents a simplified matter model, it allows for a transparent analysis due to the conservation of the mass function, and it captures a rich set of physical possibilities.
Collapse models with vanishing radial pressure were originally studied by Datta in the context of the Einstein cluster, describing counter-rotating particles \cite{Datta1970, Bondi1971, JhinganMagli2000, HaradaIguchiNakao1998}. These models have since been explored extensively to characterize collapse end states in terms of black holes and naked singularities \cite{Magli1997, Magli1998, HaradaNakaoIguchi1999, JoshiGoswami2002}. In \cite{Joshi:2011zm}, the general formalism developed in \cite{MalafarinaJoshi2011a, JoshiMalafarina2011} was employed to analyze the dynamical evolution of such systems and to determine the conditions under which collapse may halt and lead to static configurations. It was shown that both regular and centrally singular equilibrium spacetimes can arise from this process. As a specific application, they examined a toy model called JMN-1, which exhibits a central naked singularity, and studied the properties of an accretion disk in the resulting spacetime.
A notable feature of the solutions presented there is that naked singularities emerge from a fully dynamical collapse process starting from regular initial conditions.

\textcolor{black}{Recently, general relativistic magnetohydrodynamic (GRMHD) simulations of the JMN-1 naked singularity spacetime have shown that, despite the centrifugal barrier that suppresses direct accretion of rotating matter, JMN-1 can produce jet powers comparable to black holes through efficient gravitational-energy release, fluid pressure gradients, and magnetic forces \cite{Uniyal:2025hik}. Such powerful outflows may have important implications for galaxy evolution. Further, \cite{Trivedi:2025ukh} proposed that JMN-type naked singularities can provide a viable framework for understanding AGN feedback and the quenching of star formation, establishing their potential astrophysical relevance.
This connection provides the physical motivation for our study. Galaxy and AGN formation originate from the gravitational collapse of overdense regions in the universe, whose initial perturbations are encoded in the cosmic microwave background \cite{2010}. The subsequent collapse of these overdense regions depends crucially on the matter distribution within such regions\cite{2010}. The JMN collapse scenario assumes that this overdense cloud is initially homogeneous and evolves toward an equilibrium configuration described by the JMN metric. 
However, such an assumption is highly idealized. Even within an otherwise homogeneous cosmological background, a realistic overdense region is expected to possess a smooth density profile with the density higher near the center, and gradually decreasing toward the boundary. Therefore, a physically realistic generalization of the JMN model should incorporate such an inhomogeneous initial density distribution.}

\textcolor{black}{In this work, we generalize the JMN-1 model by introducing an initially inhomogeneous density profile, and we investigate its observational consequences. Further, we also test the robustness of the physical properties of the JMN-1 spacetime under this generalization. We find that, although its qualitative features remain intact, observable properties—particularly those associated with accretion-disk structure—can exhibit certain quantitative deviations.}

This paper is organized as follows. In Section~\ref{sec2}, we review the general formalism describing the dynamical evolution of matter clouds supported by tangential pressure and the conditions under which collapse asymptotes to equilibrium. Section~\ref{sec3} examines a representative generalized toy model and its physical properties. In Section~\ref{sec4}, we analyze shadow properties of this new model. In Section~\ref{sec5}, we analyze the accretion disk properties in order to distinguish the new model from the old toy model JMN-1. Finally, Section~\ref{sec6} summarizes the main results and outlines directions for future work.

\section{Equilibrium configuration}\label{sec2}
The spacetime geometry of a spherically symmetric, dynamical gravitational collapse is represented by
\begin{equation}\label{metric}
ds^2 = -e^{2\nu}dt^2 + \frac{R'^2}{G}dr^2 + R^2 d\Omega^2 ; ,
\end{equation}
where $\nu$, $R$, and $G$ are functions of the comoving time $t$ and the comoving (Lagrangian) radial coordinate $r$. The Einstein tensor for this metric reads:

\begin{equation}\label{G00}
G^{0}_{0}=-e^{-2\nu}\Big( e^{2\nu}-e^{2\nu}G -\frac{R\dot R\dot G}{G} + \dot R^{2}+\frac{ R\big(-e^{2\nu}G' + 2\dot R\dot R' \big)}{R'}\Big)\frac{1}{R^{2}}.
\end{equation}
\begin{equation}\label{G01}
G^{0}_{1}=e^{-2\nu}\frac{(R' \dot G - 2G\nu' \dot R)}{G R}.
\end{equation}
\begin{equation}\label{relation3}
G^{1}_{1}=\frac{G\Big(\frac{R'+2R\nu'}{R'}\Big)-e^{-2\nu}{\big(e^{2\nu}+\dot R^{2}-2R\dot R\dot\nu + 2R\ddot R\big)}}{R^{2}}.
\end{equation}
\begin{multline}\label{relation4}
G^{3}_{3}=G^{2}_{2}=e^{-2\nu}\Big[4e^{2\nu}G^{3}(\,R'^{2}\nu'-R\nu' R'' + R R'(\nu'^{2}+\nu'')) - 3RR'^3\dot G^2\\+2GR'^2(2R\dot G \dot R'+R'(\dot G(\dot R-R\dot \nu)+R\ddot G))+2G^2R'(e^{2\nu}G'(R'+R\nu')\\+2R'(-\dot R\dot R'+R'(\dot R\dot\nu-\ddot R)+R(\dot \nu \dot R'-\ddot R')))\Big].
\end{multline}
In the case of vanishing radial pressure, the energy-momentum tensor is given by $T^0_0=\rho, \; T_1^1=0, \; T_2^2=T_3^3=p_\theta$. Putting $G^{0}_{1}=0$, 
\begin{equation}\label{gdot}
    \dot G = 2G \dot R \frac{\nu'}{R'}
\end{equation}
Simplifying Eq.(\ref{G00}) with $\dot G$ from Eq.(\ref{gdot}), the Einstein equation gives,
\begin{equation}\label{density}
    \rho = \frac{F'}{R^2R'}
\end{equation}
where,
\begin{equation}\label{f}
    F=R(1-G+e^{-2\nu}\dot R^2),
\end{equation}
which represent the Misner-Sharp mass, which describes the amount of matter enclosed by the shell labeled by $r$. Using Eqs.(\ref{f}), (\ref{gdot}), \& (\ref{relation3}), we get,
\begin{equation}
    P_r=\frac{-\dot F }{R^2 \dot R}=0,
\end{equation}
which implies that $F=F(r)$. So the mass interior to any Lagrangian radius $r$ is conserved throughout the evolution. Therefore, at all times, the metric describing the evolving cloud can be matched to an exterior Schwarzschild solution with a total mass $\mathbf{M}$ at a boundary $r=r_b$, which corresponds to a time-dependent physical radius $R_b(t)=R(r_b,t)$ \cite{matching}. Writing $G'$ in terms of $\rho$, using Eqs.(\ref{gdot}) \& (\ref{relation4}), we get,
\begin{equation}\label{ptheta}
    p_\theta=\frac{R\rho\nu'}{2R'}.
\end{equation}
There exists a freedom in defining the physical radius \(R\), allowing a suitable rescaling without loss of generality. To exploit this, we introduce a scaling function \(v(r,t)\) such that
\begin{equation}
R(r,t) = r\,v(r,t), \qquad v(r,t_i) = 1,
\end{equation}
where the second condition ensures that the physical radius initially coincides with the comoving coordinate, \(R = r\), at the onset of collapse \(t = t_i\). Throughout the analysis, we assume the absence of shell-crossing singularities—points where \(R' = 0\). Such singularities are considered weak, arising from the intersection of neighboring shells, and can generally be removed through a suitable coordinate transformation. To avoid them, we require \(R' > 0\) at all stages of the evolution. This condition further ensures that, for positive pressures, the weak energy condition is satisfied whenever \(F' > 0\) during the collapse.

The general procedure for evolving the gravitational collapse is as follows. The system involves six unknowns, namely \(\rho\), \(p_\theta\), \(\nu\), \(G\), \(F\), and \(R\), governed by four Einstein equations. Consequently, two of these functions can be freely specified. Once the initial data for all six quantities are provided at an initial epoch \(t = t_i\), along with suitable choices for the two free functions, the system becomes closed, and the Einstein equations determine the subsequent evolution at all future times.
Typically, the free functions are chosen to be the mass function \(F(r)\)—which represents the conserved mass distribution of the collapsing cloud and allows the energy density to be obtained from equation~\eqref{density}—and the tangential pressure \(p_\theta\). With these specified, the Einstein equations fully determine the evolution \cite{initial}. In the present case, the mass function is time-independent and fixed once and for all, while the tangential pressure depends on both \(r\) and \(t\) through \(v(r,t)\), i.e. \(p_\theta = p_\theta(r,v)\). Hence, a global prescription for \(p_\theta\) as a function of \(r\) and \(v\) is required to completely fix the dynamics.
As an example, one may choose \(F = M_0 r^3\), corresponding to an initially homogeneous matter cloud of constant density. It is known that for certain choices of the functional form of \(p_\theta(r,v)\), the complete gravitational collapse can end in either a black hole or a naked singularity \cite{tangential}. For other classes of \(p_\theta\), the collapse may undergo a bounce, reversing into an expansion phase \cite{bounce}. Thus, the specific choice of the tangential pressure \(p_\theta\) dictates the future evolution of the system—whether it leads to continual collapse resulting in a black hole or a naked singularity, a bounce followed by expansion, or a stable equilibrium configuration as discussed in this work.

Any chosen form of the tangential pressure \(p_\theta\) effectively specifies an equation of state for the collapsing matter, and vice versa. The relation between the energy density and the tangential pressure is given implicitly by equation~\eqref{ptheta},
\begin{eqnarray}
    k(r,v) \equiv \frac{p_\theta}{\rho} = \frac{1}{2} R \frac{\nu'}{R'} \; .
\end{eqnarray}
Thus, \(p_\theta\) need not follow a simple idealized form—such as linear or polytropic dependence on \(\rho\). The collapsing system may evolve through diverse physical regimes, from low to extremely high densities and gravitational strengths, which is reflected in \(k(r,v)\) being a general function of \(r\) and \(t\).

Using equations~\eqref{ptheta} and~\eqref{gdot}, the metric functions \(\nu\) and \(G\) can be expressed as
\begin{eqnarray}
\nu(r,t) &=& 2\int_0^r k\,\frac{R'}{R}\,d\tilde{r} + y(t), \\[4pt]
G(r,t) &=& b(r)\,e^{4\int_v^1 \frac{k}{\tilde{v}}\,d\tilde{v}} \; .
\end{eqnarray}
The integration function \(y(t)\) can be absorbed by redefining the time coordinate, while \(b(r)\) corresponds to the velocity profile of the collapsing shells (reducing to \(G = b\) in the dust limit). The remaining dynamical variable is the physical radius \(R(r,t)\), determined by the evolution equation for \(\dot{R}\) from~\eqref{f}. The interior spacetime metric then takes the form
\[
ds^2 = -e^{4\int_0^r k\,\frac{R'}{R}\,d\tilde{r}} dt^2 
+ \frac{R'^2}{b(r)e^{4\int_v^1 \frac{k}{\tilde{v}}\,d\tilde{v}}} dr^2 
+ R^2 d\Omega^2 \; .
\]

While explicit analytic solutions for such configurations are known only in special cases (e.g., the Einstein cluster), obtaining them in full generality is difficult and often unnecessary. Assuming regular initial data and smooth free functions, one can still extract valuable information about the qualitative behaviour and structure of the dynamical collapse without performing complete integrations.

We now ask: for what forms of the tangential pressure \(p_\theta(r,v)\) (or equivalently \(p_\theta(r,t)\)) can a collapsing matter cloud, starting from regular initial data, asymptotically settle into a static equilibrium configuration? By suitably choosing \(p_\theta\), it is possible to balance gravity at late times and obtain equilibrium states consistent with physical requirements such as positive energy density and regularity of the initial data.

For any shell labeled by \(r\), the collapse dynamics can be described in terms of an effective potential,
\begin{eqnarray}
    V(r,v) = -\dot{v}^2 = -e^{2\nu}\!\left(\frac{M}{v} + \frac{G-1}{r^2}\right),
\end{eqnarray}

where \(F = r^3 M(r)\). Static or bouncing configurations require nonzero pressures—dust collapse (\(p_\theta = 0\)) always leads to a singular end state. To reach equilibrium, both velocity and acceleration must vanish:
\begin{eqnarray}
    \dot{v} = \ddot{v} = 0,
\end{eqnarray}
which corresponds to \(V = V_{,v} = 0\). These conditions define the equilibrium radius \(v_e(r)\), approached asymptotically as \(t \to \infty\).

At equilibrium, the metric functions satisfy
\begin{eqnarray}
    G_e = 1 - \frac{r^2 M}{v_e}, \qquad (G_{,v})_e = \frac{M r^2}{v_e^2},
\end{eqnarray}
and the corresponding density and tangential pressure are
\begin{eqnarray}
    \rho_e = \frac{3M + rM'}{v_e^2 (v_e + r v_e')}, \qquad 
p_{\theta e} = \frac{1}{4}\frac{r^2 M (3M + rM')}{v_e^2 (v_e + r v_e')(v_e - r^2 M)}.
\end{eqnarray}
Once \(F(r)\) and \(v_e(r)\) are specified, all equilibrium quantities are determined. The pressure function \(p_\theta(r,v)\) must then evolve such that \(p_\theta \to p_{\theta e}\) as \(t \to \infty\).

The resulting equilibrium metric is
\begin{eqnarray}\label{equi}
    ds^2_e = -e^{4\int \frac{p_{\theta e} R'_e}{\rho_e R_e} d\tilde{r}} dt^2
+ \frac{R_e'^2 (\rho_e + 4p_{\theta e})}{\rho_e} dr^2
+ R_e^2 d\Omega^2,
\end{eqnarray}
with \(R_e = r v_e\). If \(v_e(0) = 0\), the equilibrium contains a central curvature singularity, which is reached only in the infinite-time limit, representing a naked singularity, since \(F/R_e < 1\) ensures the absence of trapped surfaces.

These equilibrium solutions correspond to the class of static tangential-pressure metrics originally described by Florides~\cite{florides}, which can be obtained here as the asymptotic limit of dynamical collapse. Physically, as the collapse proceeds, the velocity and acceleration of infalling shells diminish, and the system freezes into a static configuration. Thus, the family of pressures \(p_\theta(r,v)\) leading to such limits represents a broad class of gravitational collapse models asymptoting to static equilibrium geometries.

\section{Generalization of JMN1 models}\label{sec3}
In the original JMN model \cite{Joshi:2011zm}, a simple toy model is considered with the choice
\begin{equation}
F(r)=M_0 r^3, \qquad v_e(r)=r^2, \qquad F(R)=M_0R ,
\label{toymodel}
\end{equation}
which corresponds to a collapse scenario with an initially homogeneous density profile. 
In the present work, we generalize the mass function to the form
\begin{equation}
F(r) = (M_0 + M_n r^n) r^3=M_0 R + M_n R^{(n+3)/3} ,
\end{equation}
where $M_0>0$ and $M_n<0$, while keeping $v_e(r)=r^2$. 
In principle, a more general extension could be written as, 
$F(r) = \left(\sum_{i=0}^{n} M_i r^i\right) r^3$,
which would allow for a richer class of mass distributions. However, such a choice would render the metric functions analytically intractable, leading to a purely numerical spacetime. Therefore, in this work, we restrict ourselves to the analytically tractable generalization $F(r)=(M_0+M_n r^n)r^3$.

The motivation for this generalization is physical. In the JMN toy model~\cite{Joshi:2011zm}, the initial density of the collapsing cloud is homogeneous. However, in more realistic scenarios with non-vanishing pressures, gravitational collapse need not end in a singular state, but can instead lead to equilibrium configurations. This is consistent with the formation of many astrophysical systems—such as stars and galaxies—which arise from collapse and settle into equilibrium. It is therefore of interest to examine whether such equilibrium spacetimes can emerge from more general, physically realistic initial density profiles.

\begin{figure*}[h!]
  \centering
  \begin{subfigure}[b]{0.45\textwidth}
    \includegraphics[width=\textwidth]{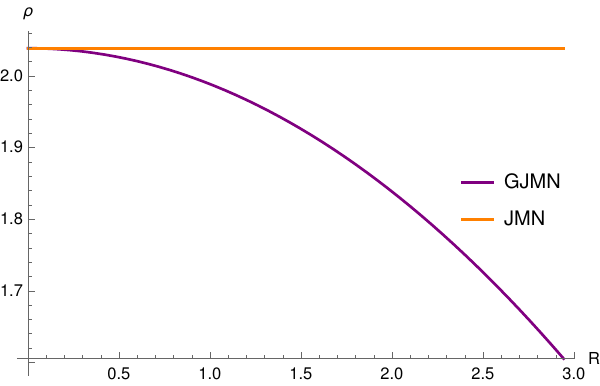}
    \caption{Initial density profile for the generalized JMN model with $M_0=0.7$ and $M_n=-0.01$, shown for the same matching radius $R_b$ as in the corresponding JMN case.}
  \end{subfigure}
  \hfill
  \begin{subfigure}[b]{0.45\textwidth}
    \includegraphics[width=\textwidth]{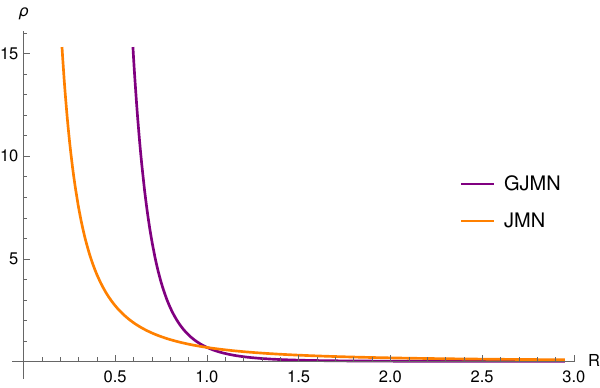}
    \caption{Corresponding equilibrium density distribution after collapse for $M_0=0.7$ and $M_n=-0.01$.}
  \end{subfigure}

  \vspace{0.5cm}

  \begin{subfigure}[b]{0.45\textwidth}
    \includegraphics[width=\textwidth]{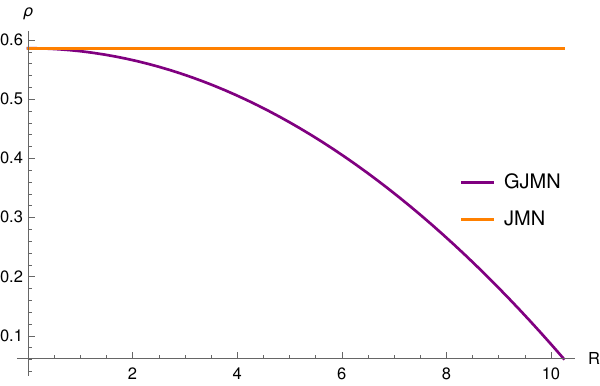}
    \caption{Initial density profile for another representative parameter choice $M_0=0.2$ and $M_n=-0.001$, again keeping $R_b$ fixed to match the JMN case.}
  \end{subfigure}
  \hfill
  \begin{subfigure}[b]{0.45\textwidth}
    \includegraphics[width=\textwidth]{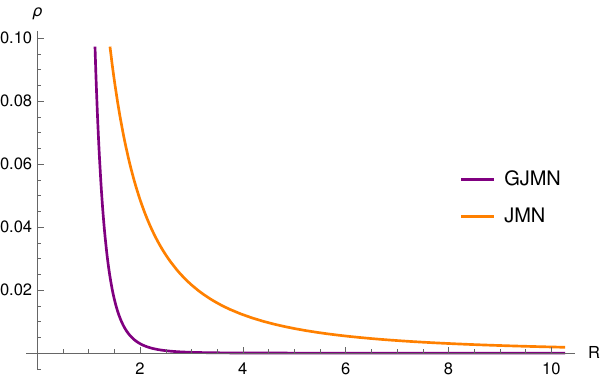}
    \caption{Corresponding equilibrium density distribution after collapse for $M_0=0.2$ and $M_n=-0.001$.}
  \end{subfigure}

  \caption{Density profiles before and after collapse for the generalized JMN (GJMN) spacetime. In all cases, the matching radius $R_b$ is kept the same as in the corresponding JMN configuration to enable a direct comparison. The additional parameter $M_n$ introduces radial inhomogeneity in the density profile, leading to equilibrium configurations that differ from the homogeneous-density JMN toy model. We have fixed $n=2$ and $\mathbf{M}=1$}
  \label{density distribution}
\end{figure*}

In Ref.~\cite{Joshi:2011zm}, collapse with vanishing radial pressure and non-zero tangential pressure was shown to produce a range of equilibrium end states, albeit with a homogeneous initial density profile (see Fig.~\ref{density distribution}). In contrast, our generalized model allows for an initial inhomogeneous density distribution.

Figure~\ref{density distribution} shows the density profiles before and after collapse for representative values of $(M_0, M_n)$. To enable a direct comparison with the JMN spacetime, we fix the matching radius $R_b$ in both the JMN and generalized JMN (GJMN) cases, ensuring that differences arise solely from modifications to the mass function.

The generalized model exhibits qualitatively different density structures. In particular, the parameter $M_n$ introduces radial inhomogeneity, and varying $(M_0, M_n)$ leads to a broader class of equilibrium configurations emerging from gravitational collapse.

\begin{figure}[h!]
    \centering
    \includegraphics[width=\linewidth]{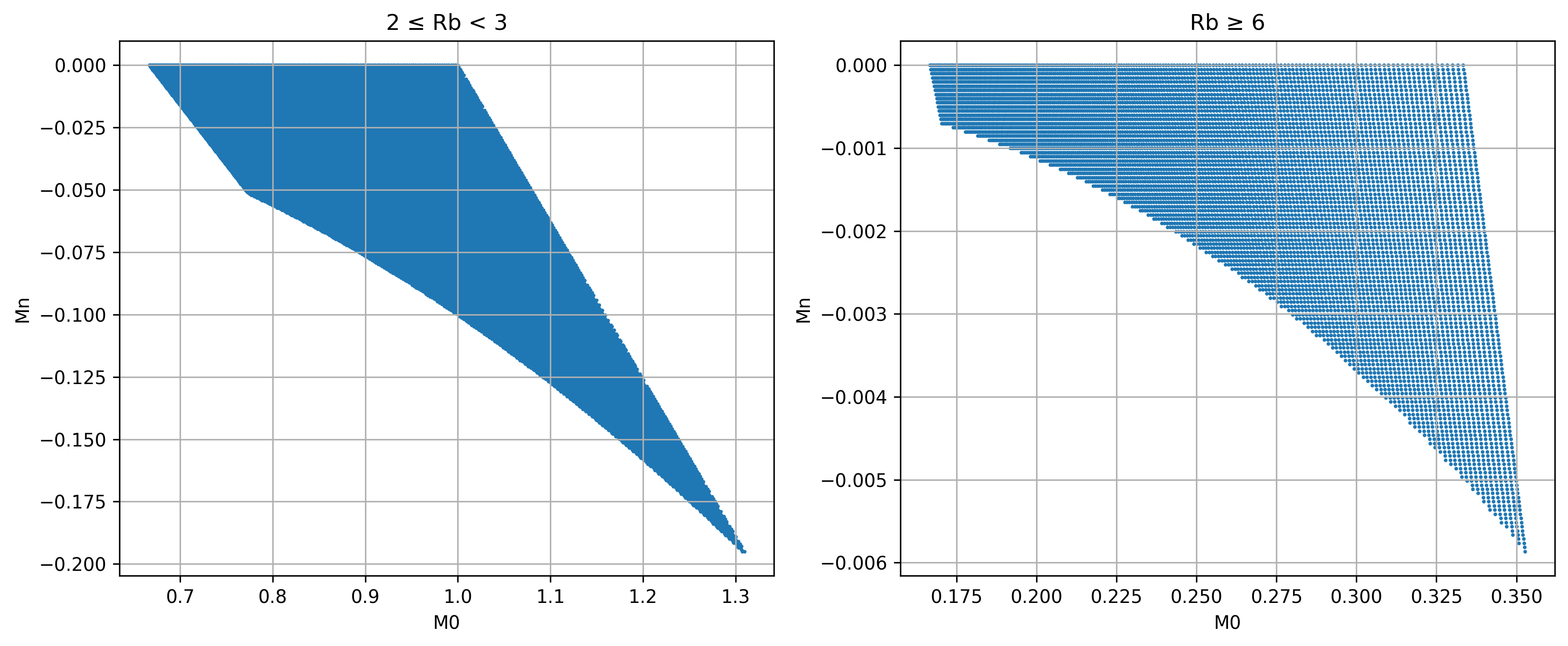}
    \caption{Allowed parameter space in the $(M_0,M_n)$ plane for two physically 
    relevant ranges of the matching radius $R_b$. The left panel corresponds 
    to $2 \le R_b < 3$, where the parameters form a continuous region. The 
    right panel corresponds to $R_b \ge 6$, where the allowed parameter space 
    collapses into thin discrete bands due to the rapid shrinking of the 
    admissible range of $M_n$ as $R_b$ increases. The plots are generated for 
    $n=2$ and $\mathbf{M}=1$.}
    \label{range}
\end{figure}

The integral in Eq.(\ref{equi}),
\begin{eqnarray}
 \int \frac{p_{\theta e} R'_e}{\rho_e R_e} d\tilde{r}=\frac{1}{4}ln\left(\frac{c\,r^{nM_0}}{-1+M_0+M_nr^n}\right)^{\frac{3}{n(1-M_0)}},
\end{eqnarray}
where $c$ is the integration constant that can be evaluated from the boundary condition. After matching at the boundary $R=R_b$ with the Schwarzschild metric, and using the Misner-Sharp mass $2\mathbf{M}=M_0R_b+M_nR_b^{(n+3)/3}$ at the boundary, we get the complete solution for the metric in the interior $R<R_b$,
\begin{eqnarray}
    ds_e^2=-\left(\frac{(1-M_0-M_nR_b^{\frac{n}{3}})^{\frac{3+n(1-M_0)}{3}}}{1-M_0-M_nR^{\frac{n}{3}}}\left(\frac{R}{R_b}\right)^{\frac{nM_0}{3}}\right)^{\frac{3}{n(1-M_0)}}dt^2+\frac{1}{1-M_0-M_nR^{\frac{n}{3}}}dR^2+R^2d\Omega^2.
\end{eqnarray}
This metric will further be addressed as GJMN. This metric matches smoothly to a Schwarzschild spacetime in the exterior $R>R_b$,
\begin{eqnarray}
    ds^2=-\left(1-\frac{M_0R_b+M_nR_b^{(n+3)/3}}{R}\right)dt^2+\left(1-\frac{M_0R_b+M_nR_b^{(n+3)/3}}{R}\right)^{-1}dR^2+R^2d\Omega^2.
\end{eqnarray}
In this model, the condition to avoid the event horizon is $M_0 + M_n R^{n/3} < 1,$ Furthermore, to satisfy the weak energy condition, we must have \(k = {p_{\theta_e}}/{\rho_e} \geq -1\), which corresponds to $M_0 + M_n R^{n/3} \leq \frac{4}{3}$ and the positivity of the energy-density is ensured with $3M_0+M_n(3+n)R^{n/3}>0$.
The effective sound speed \(c_\theta\) is given by 
$c_\theta^2 = k$,
and if we want this to be less than unity, then we require $M_0 + M_n R^{n/3} < \frac{4}{5}$.
For application purposes, once the total mass of the system is determined, we can determine the boundary of the matter distribution from the Misner-Sharp mass. We thus obtain a two-parameter family of static equilibrium configurations, characterized by \(M_0\) and \(M_n\). 
All these conditions can be combined to obtain the following bounds on the 
parameter $M_n$,
\begin{equation}
    \frac{-6\mathbf{M}}{(3+n)R_b^{n+1}-3R_b^{\frac{n+3}{3}}} < M_n < 0 ,
\end{equation}
which in turn implies the corresponding allowed range for $M_0$,
\begin{equation}
    \frac{2\mathbf{M}}{R_b} < M_0 < \frac{2\mathbf{M}}{R_b} 
    + \frac{6\mathbf{M}R_b^{\frac{n}{3}}}{(3+n)R_b^{n+1}-3R_b^{\frac{n+3}{3}}} .
\end{equation}
We focus on two physically relevant regimes of the matching radius $R_b$. 
The first regime corresponds to $R_b < 3\mathbf{M}$, where the photon sphere 
lies in the exterior Schwarzschild region. The second regime corresponds to 
$R_b > 6\mathbf{M}$, in which the outer boundary of the interior spacetime lies outside the innermost stable circular orbit (ISCO) of the Schwarzschild geometry, $r_{\mathrm{ISCO}} = 6\mathbf{M}$. In this case, the accretion disk can extend continuously from large radii all the way down to the singularity without any interruption.
For these two regimes of $R_b$, the allowed parameter space in $(M_0, M_n)$ 
is shown in Fig.~\ref{range}. In the case $2 \le R_b < 3$, the allowed values 
of the parameters form a continuous region in the $(M_0,M_n)$ plane. In 
contrast, for $R_b \ge 6$ the parameter space appears as thin discrete bands 
rather than a filled region. This behaviour arises because, as $R_b$ becomes 
large, the allowed range of $M_n$ shrinks rapidly toward zero. For the plots 
shown in the figure, we have fixed $n=2$ and $\mathbf{M}=1$.

Each solution in this family possesses a central naked singularity. These equilibrium states emerge as the final outcome of dynamical gravitational collapse starting from regular initial conditions defined by \(F(r) = M_0 r^3 + M_n r^{n+3}\), where the collapse evolution function \(v(r,t)\) approaches asymptotically the equilibrium profile \(v_e(r) \propto r^2\) as \(t \to \infty\).

In order to specify the nature of the central singularity, we note that the outgoing radial null geodesics in the spacetime above are given by,
\begin{eqnarray}
    \frac{dR}{dt}= \left(\frac{(1-M_0-M_n R_b^\frac{n}{3})^\frac{3+n(1-M_0)}{3}}{(1-M_0-M_n R^\frac{n}{3})^\frac{1-nM_0}{3}}\left(\frac{R}{R_b}\right)^\frac{nM_0}{3}\right)^\frac{-3}{2n(1-M_0)}.
\end{eqnarray}
It can be verified that outgoing radial light rays escape from the singularity for all values of $M_0 < \tfrac{2}{3}$. Moreover, one finds that the comoving time taken by a photon to reach the boundary of the collapsing cloud remains finite. Consequently, there exist future-directed null geodesics that originate at the singularity and reach the cloud boundary, thereby establishing the singularity as a naked singularity.
Along these null trajectories, both the energy density and the spacetime curvature diverge as one approaches the singularity in the past, confirming that it corresponds to a curvature singularity. The Kretschmann scalar for this naked singularity model is given by:
\begin{multline}
K=\frac{1}{36 R^4 (-1 + M_0 + M_n R^{n/3})^2}
\Big(
297 M_0^4
-18 M_0^3 (30 + M_n (-66 + n) R^{n/3})
+3 M_0^2 \Big(84 + 4 M_n (-135 + 4 n) R^{n/3}
\\+3 M_n^2 (198 - 6 n + n^2) R^{2 n/3}\Big)
+ M_n^2 R^{2 n/3}
\Big(
-6 n (4 - 8 M_n R^{n/3} + 3 M_n^2 R^{2 n/3}) \\
+ n^2 (12 - 20 M_n R^{n/3} + 9 M_n^2 R^{2 n/3})
+ 9 (28 - 60 M_n R^{n/3} + 33 M_n^2 R^{2 n/3})
\Big) \\
+ 2 M_0 M_n R^{n/3}
\Big(
M_n n^2 R^{n/3} (-10 + 9 M_n R^{n/3})
- 3 n (4 - 16 M_n R^{n/3} + 9 M_n^2 R^{2 n/3}) \\
+ 18 (14 - 45 M_n R^{n/3} + 33 M_n^2 R^{2 n/3})
\Big)
\Big).
\end{multline}
We observe that the Kretschmann scalar diverges in the limit as the approach to the central singularity is taken. The spacetime is regular everywhere for all values of $r>0$.

\section{Shadow Formation}\label{sec4}
Here, we first discuss the general procedure to obtain the shadow of a compact object. A general static, spherically symmetric geometry may be expressed as
\begin{equation}
    ds^{2} = -A(r)\,dt^{2} + B(r)\,dr^{2} + r^{2}\left(d\vartheta^{2} + \sin^{2}\!\vartheta\, d\varphi^{2}\right),
    \label{metric}
\end{equation}
where the metric functions $A(r)$ and $B(r)$ depend only on the radial coordinate, reflecting spherical symmetry.
For null geodesics confined to the equatorial plane ($\vartheta=\pi/2$), one obtains
\begin{equation}
    \frac{1}{\beta^{2}}
    = A(r)B(r)\left(\frac{dr}{d\lambda}\right)^{2}\frac{1}{L^{2}}
    + U_{\mathrm{eff}}(r),
\end{equation}
where the effective potential is $U_{\mathrm{eff}}(r)=A(r)/r^{2}$, the impact parameter is $\beta = L/E$, and $E$, $L$ are the conserved energy and angular momentum per unit mass. The relation follows from the null condition $k_{\mu}k^{\mu}=0$.
The photon dynamics are governed by $U_{\mathrm{eff}}(r)$. An unstable circular photon orbit occurs when
\begin{equation}
    U_{\mathrm{eff}}(r_{\mathrm{ph}})=\frac{E^{2}}{L^{2}},\quad 
    U'_{\mathrm{eff}}(r_{\mathrm{ph}})=0,\quad
    U''_{\mathrm{eff}}(r_{\mathrm{ph}})<0,
\end{equation}
defining the photon-sphere radius $r_{\mathrm{ph}}$. In the Schwarzschild geometry, this gives $r_{\mathrm{ph}}=3M$.
A turning point $r_{\mathrm{tp}}$ for a photon incoming from infinity satisfies  
\begin{equation}
    U_{\mathrm{eff}}(r_{\mathrm{tp}})=\frac{1}{\beta_{\mathrm{tp}}^{2}},
\end{equation}
from which the corresponding impact parameter is
\begin{equation}
    \beta_{\mathrm{tp}}=\frac{r_{\mathrm{tp}}}{\sqrt{A(r_{\mathrm{tp}})}}.
\end{equation}
If the effective potential possesses a single maximum, the minimum impact parameter for which a photon can escape to infinity is $\beta_{\mathrm{ph}}$ (the one associated with $r_{\mathrm{ph}}$). Geodesics with $\beta<\beta_{\mathrm{ph}}$ are captured, producing a shadow of radius $\beta_{\mathrm{ph}}$ on the observer’s sky. When no photon sphere exists and $U_{\mathrm{eff}}$ diverges at the center, no such shadow forms.

\begin{figure}
    \centering
    \includegraphics[width=0.5\linewidth]{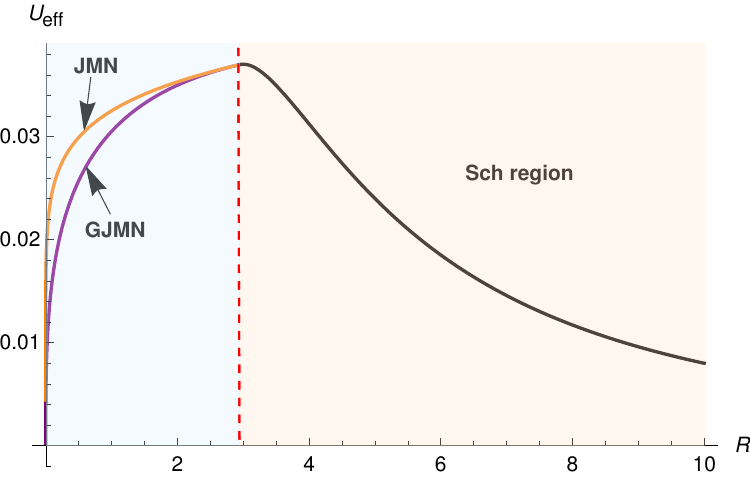}
    \caption{The effective potential for JMN1($M_0=0.6794$) and GJMN($n=2,\, M_0=0.7,\, \&\, M_2=-0.01$) spacetime inside the boundary $R_b=2.9435$, which smoothly matches with the effective potential for the Schwarzschild spacetime($\mathbf{M}=1$).}
    \label{fig:potential}
\end{figure}

In the JMN1 and GJMN spacetime, a photon sphere exists only when $R_b<3\mathbf{M}$, where $\mathbf{M}$ denotes the Schwarzschild mass; for $R_b>3\mathbf{M}$, no photon sphere is present. When $R_b<3\mathbf{M}$, the JMN1 and GJMN interior joins smoothly to an exterior Schwarzschild geometry at a matching radius $R=R_b$. Consequently, the effective photon sphere lies in the exterior Schwarzschild region, although the central singularity remains naked. The photon-sphere radius of the exterior Schwarzschild region is $r_{\mathrm{ph}}=3$. Thus, for $R_b<3\mathbf{M}$, the resulting shadow is effectively generated by the Schwarzschild photon sphere.

For comparison, Figure~\ref{fig:potential} displays the effective potential for JMN1($M_0=0.6794$) and GJMN($M_0=0.7,\, M_2=-0.01$) spacetime, both inside the boundary $R_b=2.9435$, which smoothly matches with the effective potential for the Schwarzschild spacetime.
The critical impact parameter associated with the photon sphere is $b_{\mathrm{ph}}= 3\sqrt{3}\, \mathbf{M}= 3\sqrt{3},$
which represents the shadow radius as observed at infinity.

\begin{figure*}[htbp]
  \centering
  \begin{subfigure}[b]{0.5\textwidth}
    \includegraphics[width=\textwidth]{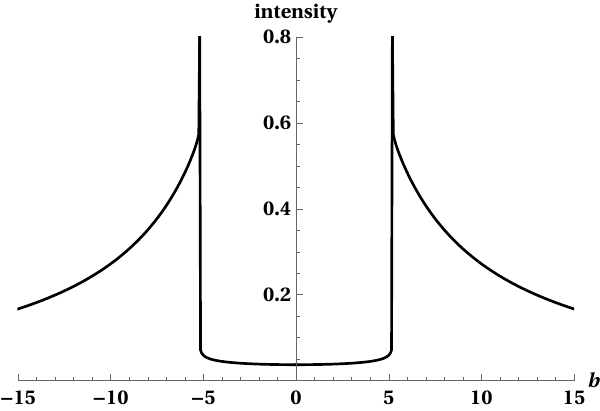} 
    \caption{Intensity distribution in GJMN spacetime with $M_0=0.7$, $M_2=-0.01$.}
    \label{fig:IGJMN}
  \end{subfigure}
  \hfill
  \begin{subfigure}[b]{0.40\textwidth}
    \includegraphics[width=\textwidth]{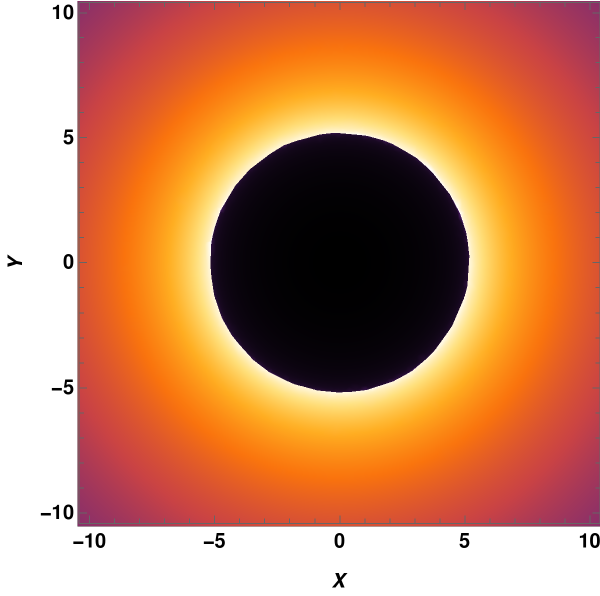} 
    \caption{Shadow in GJMN spacetime with $M_0=0.7$, $M_2=-0.01$.}
    \label{fig:SGJMN}
\end{subfigure}
\caption{Intensity distribution and the shadow cast by GJMN spacetime. We have fixed $n=2$ and $\mathbf{M}=1$}
\end{figure*}

For clarity, we model the accreting matter as a radially infalling, spherically symmetric, optically thin flow emitting monochromatic radiation with emissivity
\begin{equation}
    j(\nu_{\mathrm{e}})\propto\frac{\delta(\nu_{\mathrm{e}}-\nu_{0})}{r^{2}},
\end{equation}
where $\nu_{\mathrm{e}}$ is the photon frequency in the emitter's rest frame.
The observed specific intensity at sky coordinates $(X,Y)$ is
\begin{equation}
    I_{\nu_{\mathrm{o}}}(X,Y)=\int_{\gamma} g^{3}\, j(\nu_{\mathrm{e}})\, dl_{\mathrm{prop}},
\end{equation}
with the redshift factor
\begin{equation}
    g=\frac{\nu_{\mathrm{o}}}{\nu_{\mathrm{e}}}
      =\frac{k_{\mu}u_{\mathrm{o}}^{\mu}}{k_{\nu}u_{\mathrm{e}}^{\nu}}.
\end{equation}
For radially free-falling emitters in the metric~\eqref{metric}, the four-velocity components are
\begin{equation}
    u^{t}_{\mathrm{e}}=\frac{1}{A(r)},\qquad
    u^{r}_{\mathrm{e}}=-\sqrt{\frac{1-A(r)}{A(r)B(r)}},\qquad
    u^{\vartheta}_{\mathrm{e}}=u^{\varphi}_{\mathrm{e}}=0.
\end{equation}
The redshift factor becomes
\begin{equation}
    g=\left[
        \frac{1}{A(r)}
        -\frac{k_{r}}{k_{t}}
         \sqrt{\frac{1-A(r)}{A(r)B(r)}}
    \right]^{-1},
\end{equation}
where
\begin{equation}
    \frac{k^{r}}{k^{t}}
    =\sqrt{\frac{A(r)}{B(r)}
    \left(1-\frac{A(r)\beta^{2}}{r^{2}}\right)}.
\end{equation}
Thus, the observed intensity simplifies to
\begin{equation}
    I_{\mathrm{o}}(X,Y)\propto
    -\int_{\gamma}\frac{g^{3}k_{t}}{r^{2}k^{r}}\, dr,
    \label{intensity}
\end{equation}
 Using the eq.~(\ref{intensity}), one can simulate the shadow. In figs.~ (\ref{fig:IGJMN}), we show how intensity varies with the impact parameter $\beta$ in the GJMN naked singularity spacetime. In figs.~ (\ref{fig:SGJMN}), we simulate the shadows cast by the GJMN naked singularity spacetime.

From fig.~ ~(\ref{fig:SGJMN}), one can see that due to the presence of a photon sphere in the external Schwarzschild spacetime of the GJMN spacetime configuration (with $R_b<3\mathbf{M}$), the shadow radius is $\beta_{ph}=3\sqrt3 \mathbf{M} = 3\sqrt3$. Hence, the shadow cast by GJMN is similar to the JMN1\cite{rajbul1} spacetime configuration with $R_b<3\mathbf{M}$ and is not distinguishable from the shadow cast by a Schwarzschild black hole. Otherwise, GJMN will also cast the full moon-like shadow as given for JMN in \cite{rajbul1}.

\section{Accretion disk properties}\label{sec5}
Consider a test particle moving within the matter \lq cloud\rq~of a spherical configuration. The particle is assumed not to interact with the material constituents of the cloud and is influenced solely by the gravitational field generated by the cloud. For convenience, the total gravitational mass of the cloud is normalized to unity. Owing to the spherical symmetry of the spacetime, the coordinate can be chosen as $\theta=\pi/2$, ensuring that the geodesic motion of the test particle lies entirely in the equatorial plane. For the metric given in Eqn.(\ref{metric}) circular geodesic motion, the energy per unit mass $E$, angular momentum per unit mass $L$, and angular velocity $\omega$ of the test particle are given by~\cite{RN_1304}

\begin{equation}
E^2=\frac{2A^2}{2A-A_{,r}r},
\label{E_accretion}
\end{equation}

\begin{equation}
\frac{L^2}{r^2}=\frac{A_{,r}r}{2A-A_{,r}r},
\label{L_accretion}
\end{equation}

\begin{equation}
\omega^2=-\frac{g_{tt,r}}{g_{\varphi\varphi,r}}=\frac{A_{,r}}{2r}.
\label{omega_angular_frequency}
\end{equation}

In our analysis, we restrict ourselves to the scenario where the outer boundary of the sphere, $R_b$, lies outside the innermost stable circular orbit (ISCO), located at $r_{\scriptsize\mbox{ISCO}}=6M_{T}$. Under this condition, the accretion disk extends smoothly from large radii down to the singularity without any discontinuity~\cite{RN_1304}. The radiative properties of such disks can be evaluated using the formalism developed in Refs.~\cite{Novikov,Page}. As emphasized in Ref.~\cite{RN_1304}, the spectral luminosity distribution provides a useful diagnostic tool for distinguishing different density profile models from the Schwarzschild black hole scenario.

We assume that each infinitesimal element of the accretion disk emits radiation as a blackbody. A characteristic temperature $T_*$ is defined through the relation $\sigma T_*^4 \equiv \dot{m}c^2/[4\pi(GM_T/c^2)^2]$, where $\dot{m}$ denotes the constant rest mass accretion rate and $\sigma$ is the Stefan--Boltzmann constant. Let $\mathcal{F}(r)$ represent the radiative flux, i.e., the energy emitted per unit area per unit time in the local rest frame of the accreting fluid (see Eq.~(\ref{flux})). The local blackbody temperature is therefore given by $T_{\rm BB}(r)=[{\mathcal F}(r)]^{1/4}T_*$.

The emitted radiation experiences both gravitational and Doppler redshifts before reaching a distant observer. The magnitude of this redshift depends on the orientation of the observer relative to the disk axis. For simplicity, we consider an observer located along the axis of the disk. The corresponding redshift factor is
\[
1+z(r)=[-(g_{tt}+\omega^2 g_{\varphi\varphi})]^{-1/2}.
\]

Assuming isotropic emission, the radiation emitted at radius $r$ corresponds to a temperature measured at infinity of $T_\infty(r)=T_{\rm BB}(r)/(1+z)$. Consequently, the spectral luminosity distribution ${\cal L}_{\nu,\infty}$ observed by a face-on observer at infinity can be approximated as

\begin{equation}
\nu {\cal L}_{\nu,\infty} = \frac{15}{\pi^4} \int_{r_{\rm inner}}^\infty \left(\frac{d{\cal L}_\infty}{d\ln r}\right)\frac{(1+z)^4 (h\nu/kT_*)^4/\mathcal{F}}
{\exp[(1+z)(h\nu/kT_*)/{\mathcal F}^{1/4}]-1}\, d\ln r,
\label{integral_L}
\end{equation}

where

\begin{equation}
\frac{d{\cal L}_\infty}{d\ln r} = 4\pi r \sqrt{-g} E {\mathcal F},
\label{dLdr}
\end{equation}

\begin{equation}
\mathcal{F}(r)=-\frac{\dot{m}}{4\pi \sqrt{-g}}\frac{\omega_{,r}}
{(E-\omega L)^2}\int^r_{r_{\rm inner}}(E-\omega L)L_{,{\tilde{r}}}d{\tilde{r}}.
\label{flux}
\end{equation}

Here, $g$ represents the determinant of the metric of the three-dimensional subspace $(t,r,\varphi)$. The quantity $r_{\rm inner}$ denotes the radius of the inner edge of the accretion disk, corresponding to the innermost stable circular orbit. For density profile models, we set $r_{\rm inner}=0$, whereas for the Schwarzschild black hole model, it is fixed at $6\mathbf{M}$.

\begin{figure}[h!]
    \centering
    \includegraphics[width=\linewidth]{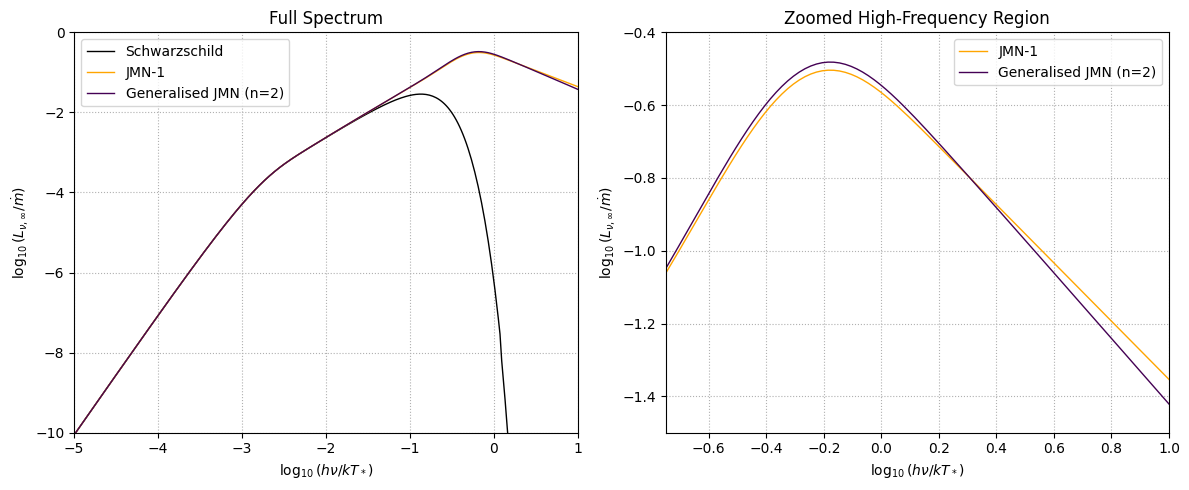}
    \caption{Accretion disk spectral luminosity distribution for the Schwarzschild, JMN, and generalized JMN (GJMN) spacetimes. The left panel shows the full spectrum, while the right panel zooms into the high–frequency region. The plots are generated for $n=2$, $\mathbf{M}=1$, $M_n=-0.005$, $M_0=0.3498$ and $R_b=6\mathbf{M}$. The same value of $R_b$ is used for both the JMN and GJMN models to allow a direct comparison.}
    \label{fig:GJMNspec}
\end{figure}

Figure~\ref{fig:GJMNspec} shows the accretion disk spectral luminosity 
for the Schwarzschild black hole, the JMN spacetime, and the generalized 
JMN (GJMN) model with $n=2$. The left panel displays the full spectrum 
over a wide frequency range, while the right panel zooms into the 
high–frequency region where small deviations between the models become 
visible.

In these calculations we focus on the regime $R_b \ge 6\mathbf{M}$ so that 
the boundary of the interior spacetime lies outside the innermost stable 
circular orbit (ISCO) of the exterior Schwarzschild geometry. In this case 
the accretion disk extends continuously from large radii all the way to the 
central singularity without any truncation. However, maintaining this regime 
while simultaneously satisfying the physical constraints of the spacetime—
such as the absence of an event horizon and the weak energy condition—
strongly restricts the allowed values of the parameter $M_n$. In particular, 
for large $R_b$ the allowed range of $M_n$ shrinks rapidly toward zero. 
Consequently, the generalized model represents only a small perturbation of 
the original JMN spacetime.

This behaviour is reflected in the spectral profiles. As seen in the left 
panel of Fig.~\ref{fig:GJMNspec}, both the JMN and the generalized JMN models 
produce significantly higher luminosity at high frequencies compared to the 
Schwarzschild case. This enhancement arises because the accretion disk in 
these spacetimes can extend all the way to the central singularity, allowing 
matter to reach deeper gravitational potentials and emit more energetic 
radiation.

A closer inspection of the high–frequency region, shown in the right panel, 
reveals that the spectra of the JMN and GJMN models are almost identical. 
The difference between the two curves is extremely small, which is expected 
because the parameter $M_n$ must remain very small when $R_b \ge 6\mathbf{M}$ 
in order to satisfy the physical constraints of the spacetime. As a result, 
the generalized mass function introduces only a mild radial inhomogeneity in 
the interior geometry, leading to very small modifications of the accretion 
disk emission relative to the original JMN model.

\section{Discussion and Conclusions}\label{sec6}
In this work, we have investigated the robustness of the Joshi--Malafarina--Narayan (JMN) naked singularity spacetime \cite{Joshi:2011zm} by constructing and analyzing a generalized class of equilibrium configurations arising from gravitational collapse with non–vanishing tangential pressure. The original JMN model corresponds to a collapse scenario with an initially homogeneous density profile, characterized by the mass function $F(r)=M_0 r^3$. In the present work, we introduced a controlled generalization of this toy model by allowing a small radial inhomogeneity in the initial mass distribution through the generalized mass function $F(r) = (M_0 + M_n r^n) r^3$. This modification preserves the essential structure of the collapse model while introducing an additional physical parameter that encodes density gradients in the collapsing cloud.

Using the general formalism describing collapse with vanishing radial pressure and non–zero tangential pressure (Sec.~\ref{sec2}), we derived the corresponding equilibrium spacetime that emerges asymptotically from the collapse dynamics. The resulting metric, which we refer to as the generalized JMN (GJMN) spacetime, is given in Sec.~\ref{sec3} and represents a two–parameter family of equilibrium configurations characterized by $(M_0, M_n)$. The spacetime matches smoothly to an exterior Schwarzschild geometry at the boundary radius $R_b$, with the total mass determined by $2\mathbf{M}=M_0R_b+M_nR_b^{(n+3)/3}$. We then imposed the physical requirements that the spacetime must (i) avoid the formation of an event horizon, (ii) satisfy the weak energy condition, and (iii) maintain subluminal effective sound speed. These conditions restrict the allowed parameter space of $(M_0, M_n)$, which we derived analytically and illustrated in Fig.~\ref{range}. An interesting feature that emerges from this analysis is that when the matching radius $R_b$ is large, the allowed values of $M_n$ become extremely small. Physically, this implies that strong density inhomogeneities are not permitted if the boundary of the matter cloud lies far outside the Schwarzschild radius. As a consequence, the generalized spacetime represents only a small perturbation of the original JMN configuration in this regime.

We next investigated the physical properties of this generalized spacetime through two key observational diagnostics: shadow formation and accretion disk emission.

First, we analyzed the null geodesic structure of the spacetime and the resulting shadow formation. As discussed in Sec.~\ref{sec4}, when $R_b<3\mathbf{M}$ the photon sphere responsible for the shadow lies in the exterior Schwarzschild region. Consequently, the shadow radius is determined entirely by the Schwarzschild photon sphere and is given by $\beta_{\mathrm{ph}} = 3\sqrt{3}\,\mathbf{M}$. Since this photon sphere lies outside the interior region, the shadow size is completely insensitive to the internal structure of the spacetime. As illustrated in Fig.~\ref{fig:SGJMN}, the shadow produced by the GJMN spacetime is therefore identical to that of the original JMN model and indistinguishable from the Schwarzschild black hole shadow. This result demonstrates the remarkable robustness of the JMN geometry with respect to perturbations in the interior density profile. Even when the collapse model is generalized to include inhomogeneities, the shadow observable remains unchanged.

Second, we studied the accretion disk properties of the generalized spacetime using the Novikov--Thorne thin disk formalism \cite{Novikov, Page}. In this analysis, we focused on the physically important regime $R_b \ge 6\mathbf{M}$, in which the boundary of the interior spacetime lies outside the Schwarzschild ISCO. In this case, the accretion disk extends continuously from large radii all the way to the singularity. The spectral luminosity distribution for the Schwarzschild, JMN, and GJMN models is shown in Fig.~\ref{fig:GJMNspec}. As expected, both the JMN and GJMN spacetimes exhibit significantly enhanced high–frequency luminosity compared to the Schwarzschild black hole. This enhancement arises because the absence of an event horizon allows the accretion disk to extend to arbitrarily small radii, enabling matter to release more gravitational energy near the central singularity. However, when comparing the JMN and GJMN spectra directly, we find that the difference between the two curves is extremely small. This behaviour can be understood from the parameter constraints discussed earlier. For $R_b \ge 6\mathbf{M}$, the allowed range of $M_n$ shrinks rapidly toward zero, forcing the generalized mass function to remain very close to the original JMN form. As a result, the generalized spacetime can effectively be interpreted as a small perturbation of the JMN geometry. The near coincidence of the two spectra therefore indicates that the JMN spacetime is \textcolor{black}{robust} under such perturbative generalizations.

Taken together, these results provide strong evidence that the JMN naked singularity spacetime is a remarkably robust solution from the observational point of view. Both the shadow structure and the accretion disk emission remain largely unchanged when small density inhomogeneities are introduced in the collapse model. This suggests that the observational properties of collapse-generated naked singularities may be relatively insensitive to moderate variations in the initial density distribution.

Finally, it is important to emphasize that the generalization studied in this work represents only the simplest extension of the JMN toy model. In principle, a much richer class of equilibrium configurations could be obtained by considering a more general mass function of the form $F(r) = \left(\sum_{i=0}^{N} M_i r^i\right) r^3 $. Such models would allow for arbitrary density profiles and could capture a wider class of physically realistic collapse scenarios. However, in that case, the resulting spacetime geometry would no longer admit simple analytic expressions for the metric functions, and the analysis would necessarily become fully numerical. A detailed investigation of such numerical equilibrium configurations, together with their observational signatures, represents an important direction for future work. In particular, it would be interesting to examine whether larger deviations in the density profile could produce observable modifications in the accretion disk emission or other astrophysical signatures while still preserving the spacetime's collapse origin.

In summary, we have constructed and analyzed a generalized class of collapse-generated naked singularity spacetimes and demonstrated that the key observational properties of the JMN model remain remarkably \textcolor{black}{robust} under small perturbations of the initial density distribution. These results reinforce the role of JMN-type geometries as important theoretical laboratories for exploring the physics of gravitational collapse and the possible observational signatures of naked 
singularities.

\vspace{1cm}
J. V. T. would like to acknowledge insightful discussions with Saurabh regarding the parameter-space plots. J. V. T. also gratefully acknowledges Ashok B. Joshi for sharing his expertise in Mathematica, particularly in relation to the construction of shadow plots.

\end{document}